\documentstyle[preprint,prl,aps]{revtex}

\draft
\newcommand{\be}{\begin{equation}}
\newcommand{\ee}{\end{equation}}
\newcommand{\ben}{\begin{eqnarray}}
\newcommand{\een}{\end{eqnarray}}

\begin{document}
\author{B. R. Frieden$^1$, A. Plastino$^{1,\,2,\,3}$, A. R. Plastino$%
^{2,\,3} $, and B. H. Soffer$^{1,\,4}$}
\address{$^1$University of Arizona at Tucson \\
$^2$ Universidad Nacional de La Plata, \\
C.C. 727, 1900 La Plata, Argentina \\
$^3$Argentine National Research Center (CONICET) \\
$^4$ Postal address: 665 Bienveneda Avenue, Pacific Palisades,\\
California 90272}
\title{A Schroedinger link between non-equilibrium thermodynamics and Fisher
information}
\maketitle

\begin{abstract}
It is known that equilibrium thermodynamics can be deduced from a
constrained Fisher information extemizing process. We show here that, more
generally, both non-equilibrium and equilibrium thermodynamics can be
obtained from such a Fisher treatment. Equilibrium thermodynamics
corresponds to the ground state solution, and non-equilibrium thermodynamics
corresponds to excited state solutions, of a Schroedinger wave equation
(SWE). That equation appears as an output of the constrained variational
process that extremizes Fisher information. \ Both \ equilibrium- and
non-equilibrium situations can thereby be tackled by one formalism that
clearly exhibits the fact that thermodynamics and quantum mechanics can both
be expressed in terms of a \ formal SWE, out of a common informational
basis. \
\end{abstract}


\bigskip \newpage

\section{Introduction}

The information content of a normalized probability distribution $%
P(i),i=1,\ldots ,N$, where the index $i$ runs over the states of the system
one is trying to study, is given by Shannon's information measure (IM) \cite%
{katz}

\begin{equation}
S=-\sum_{i=1}^{N}\,P(i)\,\ln [P(i)].  \eqnum{1}
\end{equation}%
\noindent The choice of the logarithmic base fixes the information units. If
the basis is $2$ then $S$ is measured in {\it bits }. If one chooses
Boltzmann's constant as the informational unit and identifies Shannon's IM
with the thermodynamic entropy, then the whole of statistical mechanics can
be elegantly reformulated by extremization of Shannon's $S$, subject to the
constraints imposed by the {\it a priori} information one may possess
concerning the system of interest \cite{katz}.

Now, the phenomenal success of thermodynamics and statistical physics
crucially depends upon certain necessary mathematical relationships
involving energy and entropy (Legendre transform structure). In the {\it %
equilibrium situation} these relationships are also valid if one replaces $S$
by Fisher's information measure $I$\ (FIM) \cite{nosotros}. Using this
measure \cite{fisher}, the entire Legendre-transform structure of
thermodynamics can be re-expressed (i.e., $I$ replaces Boltzmann-Shannon $S$%
). In general, this abstract Legendre structure constitutes an essential
ingredient that allows one to build up a statistical mechanics. Fisher
information $I$ allows then for such a construction. Also, a desired
concavity property, obeyed by $I$, further demonstrates its utility as a
statistical mechanics generator.

Here we will show that the variational treatment of Fisher information also
accounts for {\it non-equilibrium situations}. We will {\em connect} Fisher
information $I$ with non-equilibrium thermodynamics via the Schroedinger
equation (SWE). Such a connection is of interest because it clearly shows
that equilibrium and non-equilibrium states have a common informational
origin that is expressed by the SWE. \ \ The same SWE also allows for
quantum scenarios, or even mixed quantum and thermodynamic scenarios.

The interested reader might want to consult works by Frieden, Soffer,
Nikolov, Plastino, Silver, Hughes, Reginatto, Hall, and others, that have
shed much light upon the manifold physical applications of Fisher's
information measure \cite%
{f0,f1,f2,f3,f4,f5,f6,f7,f8,silver,pla1,pla2,pla3,pla4}. Frieden and Soffer
have shown that FIM provides one with a powerful variational principle that
yields the canonical Lagrangians of theoretical physics \cite{f7}.
Additionally, $I$ has been shown to characterize an ``arrow of time'' with
reference to the celebrated Fokker-Planck equation \cite%
{pla2,pla3,pla4,pla5,roybook}.

\section{Fisher's information measure for translation families. A
variational treatment}

Consider a system that is specified by a physical parameter $\theta $ at a
given time $t.$\ Let $g(x,\theta |t)$ describe the probability density
function (PD) for this parameter at that time. \ Of course, by normalization,%
\begin{equation}
\int dxg(x,\theta |t)=1.  \eqnum{2}
\end{equation}%
The Fisher information measure (FIM) $I$ is of the form
\begin{equation}
I\,=\,\int \,dx\,g\,\left[ \frac{\partial g/d\theta }{g}\right] ^{2},\text{
\ \ \ }g=g(x,\theta |t).  \eqnum{3}
\end{equation}

The special case of {\it translation families} is of use. These are
mono-parametric families of distributions of the form
\begin{equation}
g(x,\theta |t)=p(u|t),\text{ \ \ \ }u\equiv x-\theta ,  \eqnum{4}
\end{equation}%
which are known up to the shift parameter $\theta $. Following Mach's
principle, all members of the family possess identical shape $p(u|t)$ (there
are no absolute origins). Here FIM takes the appearance [21]
\begin{equation}
I=\int dx\frac{(\partial p/\partial x)^{2}}{p},\text{ \ \ \ }p=p(x|t).
\eqnum{5}
\end{equation}

Our present considerations assume one is dealing with coordinates ${x}$ that
belong to ${\cal R}$. Let us focus attention upon the positive-definite,
normalized PDF $p({x|t})$, evaluated at the time $t.$ It of course obeys
normalization
\begin{equation}
\int \,dx\,p(x|t)\,=\,1.  \eqnum{6}
\end{equation}

Let the mean values%
\begin{equation}
\theta _{k}\equiv <A_{k}>\text{of }M\text{ functions }A_{k}(x),\text{ \ }%
k=1,...,M  \eqnum{7}
\end{equation}%
be measured at the time $t$. By definition

\begin{equation}
\langle A_{k}\rangle _{t}\,=\,\int \,dx\,A_{k}(x)\,p(x|t),\,\,\,k=1,\dots ,M.
\eqnum{8}
\end{equation}%
These mean values will play the role of thermodynamical variables, as
explained in [2].

It is of importance to note that the prior knowledge (8) {\it represents
information at the fixed time} $t$. The problem we attack is to find the PDF
$p\equiv p_{MFI}$ that extremizes $I$ subject to prior conditions (6)-(7).
Our Fisher based extremization problem takes the form
\begin{equation}
\delta _{p}\{I(p)\,-\,\alpha \langle 1\rangle \,-\,\sum_{k}^{M}\lambda
_{k}\langle A_{k}\rangle _{t}\}\,=\,0,\text{ \ \ }p\equiv p(x|t),  \eqnum{9}
\end{equation}%
at the given time $t$. Eq. (9) is equivalent to
\begin{equation}
\delta _{p}\{\int \,dx\,\left( F_{Fisher}(p)\,-\,\alpha
f\,-\,\sum_{k}^{M}\lambda _{k}A_{k}p\right) \}\,=\,0,  \eqnum{10}
\end{equation}%
where we have introduced the ($M+1$) Lagrange multipliers ($\alpha ,\lambda
_{1}\dots \lambda _{M}$), where each Lagrange multiplier $\lambda _{k}\equiv
\lambda _{k}(t)$. Variation leads now to

\begin{equation}
\int \,dx\,\delta p\{(p)^{-2}\,(\frac{\partial p}{\partial x})^{2}\,+\frac{%
\partial }{\partial x}[(2/p)\,\frac{\partial p}{\partial x}]\,+\,\alpha
\,+\,\sum_{k}^{M}\lambda _{k}A_{k}\}\,=\,0,  \eqnum{11}
\end{equation}%
and, on account of the arbitrariness of $\delta p$%
\begin{equation}
\{(p)^{-2}\,(\frac{\partial p}{\partial x})^{2}\,+\frac{\partial }{\partial x%
}[(2/p)\,\frac{\partial p}{\partial x}]\,+\,\alpha \,+\,\sum_{k}^{M}\lambda
_{k}A_{k}\}\,=\,0.  \eqnum{12}
\end{equation}%
It is clear that the normalization condition on $p$\ makes $\alpha $ a
function of the $\lambda _{i}$'s. Let then $p_{I}(x,\{{\bf \lambda \}})$\ be
a solution of (12), where obviously, $\{{\bf \lambda \}}$ is an $M$%
-dimensioned Lagrange multipliers vector. The extreme Fisher information is
now a function of time
\begin{equation}
I=\int dx\frac{(\partial p/\partial x)^{2}}{p}\equiv I(t),  \eqnum{13}
\end{equation}%
since $p=p(x|t)$. \ Since $p$ extremized $I$ we write
\[
p\equiv p_{I},\text{ \ \ }p_{I}\equiv p_{I}(x|t).
\]

Let us now find the general solution of Eq. (12). For the sake of simplicity
let us define
\begin{equation}
G(x,t)=\alpha \,+\,\sum_{k}^{M}\lambda _{k}(t)A_{k}(x),  \eqnum{14}
\end{equation}%
and recast (12) as

\begin{equation}
\lbrack \frac{\partial \ln p_{I}}{\partial x}]^{2}\,+\,2\frac{\partial
^{2}\ln p_{I}}{\partial x^{2}}\,+\,G(x)\,=\,0.  \eqnum{15}
\end{equation}%
We introduce now the identification \cite{silver} $\ p_{I}=(\psi )^{2}$,
recalling that $\psi (x)$ can always be assumed real for one-dimensional
problems [2]. Introduce now the new functions
\begin{equation}
v=\frac{\partial \ln \psi }{\partial x},\text{ \ }\psi \equiv \psi (x,t),%
\text{ \ }v\equiv v(x,t).  \eqnum{16}
\end{equation}%
Then (15) simplifies to
\begin{equation}
v^{\prime }=-\{\frac{G}{4}+v^{2}\},  \eqnum{17}
\end{equation}%
where the prime stands for the derivative with respect to $x$. The above
equation is a Riccati equation \cite{paul}. Introduction further of \cite%
{paul}
\begin{equation}
u=\exp \{\int^{x}\,dx\,[v]\},u=u(x,t)  \eqnum{18}
\end{equation}%
i.e.,
\begin{equation}
u=\exp \{\int^{x}\,dx\,\frac{d\ln \psi }{dx}\}=\psi ,  \eqnum{19}
\end{equation}%
places (15) in the form of a Schro{e}dinger wave equation (SWE) \cite{paul}
\begin{equation}
-(1/2)\psi ^{\prime \prime }\,-\,(1/8)\sum_{k}^{M}\lambda _{k}(t)A_{k}\,\psi
\,=\,\alpha \psi /8,  \eqnum{20}
\end{equation}%
where the Lagrange multiplier $\alpha /8$ plays the role of an energy
eigenvalue, and the sum of the $\lambda _{k}A_{k}(x)$ is an effective
potential function
\begin{equation}
U=(1/8)\sum_{k}^{M}\,\lambda _{k}(t)A_{k},U=U(x,t).  \eqnum{21}
\end{equation}%
\newpage

Note that no specific potential has been assumed, as is appropriate for
thermodynamics. Also, we remark that $U$ is a time-dependent potential
function and will permit non-equilibrium solutions. The specific $A_{k}(x)$
to be used here depend upon the nature of the physical application at hand
(cf. Eq. (8)). This application could be of either a classical or a quantum
nature.

Also notice that equation (20) represents a boundary value problem,
generally with multiple solutions, in contrast with the unique solution one
obtains when employing Jaynes-Shannon's entropy in place of FIM \cite{katz}.
As discussed in some detail in \cite{nosotros}, the solution leading to the
lowest $I$-value is the equilibrium one. That was the only solution
discussed there. Here we wish to generalize the concomitant discussion and
ask: can we choose other solutions?

\bigskip

\bigskip

\section{Rumer and Ryvkin's approach to non-equilibrium thermodynamics}

\bigskip

In Ref. \cite{librobis}, Rumer and Ryvkin (R-R) use the conventional
Boltzmann transport equation to build up  non-equilibrium solutions. They
take the following approach:

\begin{itemize}
\item Consider a non-equilibrium state of a gas after the lapse of a time $t$
large compared to the time of initial randomization. The time $t$ is
regarded as $fixed$.

\item The time $t$ is, also, small compared to the macroscopic relaxation
time $T^*$ for attaining the Maxwell-Boltzmann law $f_0$ on velocities.

\item At each point of the vessel containing the gas a state arises which is
close to the {\it local} equilibrium state
\[
f_{0}=Maxwell-Boltzmann\,\,law\,\,on\,\,velocities.
\]

\item This allows one to expand the non-equilibrium distribution $f(x|t)$ as
\begin{equation}
f_{0}=1+\epsilon \phi (x,t),\text{ \ \ }f_{0}\equiv f(x,t)  \eqnum{22}
\end{equation}%
where $\epsilon $ is small and the function $\phi $ is to be the object of
our endeavors.

\item The unknown function $\phi (x,t)$ may itself be expanded as a series
of (orthogonal) Hermite-Gaussian polynomials $H_{i}(x)$ with coefficients $%
a_{i}(t)$ at the fixed time $t,$%
\begin{equation}
f(x,t)=\Sigma \text{ }a_{i}(t)H_{i}(x)\text{\thinspace }  \eqnum{23}
\end{equation}

It is important to remark that Hermite-Gaussian polynomials are orthogonal
with respect to a Gaussian kernel, i.e., the {\it equilibrium distribution}.
No other set of functions is orthogonal (and complete) with respect to a
Gaussian kernel function.

\item Because of orthogonality, the unknown coefficients $a_{i}(t)$ relate
linearly to appropriate (unknown) moments of $f$ over velocity space ($x$%
-space).

\item Substituting the expansion for $f$ into the transport equation and
integrating over all velocities yields now a set of first-order differential
equations in the moments (which are generally a function of the fixed time
value $t$).

\item These are now solvable subject to {\it known initial conditions}, like
our expectation values. The moments now become known ({\it including any
time dependence}).

\item As a consequence the coefficients $a_{i}(t)$ of (23) are also known,
which gives $f$.
\end{itemize}

What does the $f$ as determined above represent? According to Ref. \cite%
{librobis}, the solution of the above system of equations would be
equivalent to the exact solution of Boltzmann's equation (if enough a priori
information were available).

We emphasize that R-R do not use an SWE in their approach.

\bigskip

\section{Connecting the SWE excited solutions to non-equilibrium
thermodynamics}

\bigskip

Returning to our analysis, we ask: Can the excited SWE solutions to Eq. (20)
represent non-equilibrium states of thermodynamics \cite{f7,roybook}? In
order to answer this question, consider again the case in which $x$ is a
velocity and one seeks the non-equilibrium probability $p(x|t).$

Let excited solutions $\psi _{n}(x,t)$ to the SWE Eq. (20) be identified by
a subindex value $n>0.$ These amplitude functions are superpositions of
Hermite-Gaussian polynomials of the form \ \ \ \ \ \ \ \ \ \ \ \ \ \ \ \ \ \
\ \ \ \ \ \ \ \ \ \ \ \ \ \ \ \ \ \ \ \ \ \ \ \ \ \ \ \ \ \ \ \ \ \ \
\begin{equation}
\psi _{n}(x,t)=\Sigma _{i}b_{in}(t)H_{i}(x),\text{ \ \ }n=1,2,...\ \
\eqnum{24}
\end{equation}%
The total number of coefficients $b_{ni}(t)$ depends upon how far from
equilibrium we are. \ At equilibrium there is only one such coefficient.

We will show that the squares of these amplitudes agree, under certain
conditions (see below), with the known solutions of the Boltzmann transport
equation [11,21,23]. Our coefficients $b_{in}(t)$ are computed at the fixed
time $t$ at which our input data $<A_{k}>_{t}$ are collected. While the
ground state solution of (20) gives the equilibrium states of thermodynamics
[2], the excited solutions of (20) will be shown to be give non-equilibrium
states. For this to happen, our functions $\psi _{n}(x,t)$ will have to be
connected to the R-R $f(x,t)$ of Eq. (23) via the squaring operation $\psi
_{n}^{2}(x,t)$.\bigskip\

Notice that the square of an expansion in Hermite-Gaussians is likewise a
superposition of Hermite-Gaussians, with coefficients $c_{in}(t)$\bigskip

\begin{equation}
\psi _{n}^{2}(x,t)=\Sigma _{i}\text{ }c_{in}(t)H_{i}(x),\text{ \ }n=1,2,...\
\eqnum{25}
\end{equation}%
We argue now to the effect that, for fixed $n,$ the R-R coefficients $%
a_{i}(t)$ and our $c_{in}(t)$ are equal.

First of all, the R-R coefficients are certainly computed, like ours, at a
{\it fixed} time $t$. That is, their momenta are evaluated at that time. \
Likewise ours (the $<A_{k}>$ of (8)) can be regarded as velocity momenta at
that time as well.

The difference between the R-R coefficients and ours is one of physical
origin, as follows. R-R {\it solve for} the velocity moments at the fixed
time $t$. These $M_{RR}$ moments are computed using the R-R $a_{i}$ of Eq.
(23). We, instead, collect $as$ $experimental${\it \ inputs} these velocity
moments (at the fixed time $t$). Thus, if the $M_{RR}$ moments {\em coincide
with our experimental inputs}, necessarily the $a_{i}(t)$ and the $c_{in}(t)$
have to coincide well . Let us repeat: the R-R moments at the time $t$ are
physically correct by construction, since they actually solve for them via
use of the Boltzmann transport equation. The premise of our constrained
Fisher information approach is that its input constraints (here our velocity
moments $<A_{k}>_{t}$) are correct, since they come $from$ $experiment$.
(They calculate, we measure.) \

If there is no agreement between the R-R moments and our experimental
inputs, two possibilities come to mind: a) we are measuring inputs showing
strong quantum effects, while the R-R treatment can not handle such a case
(being classical), or b) the number $M$ of available experimental data we
use as inputs does not equal the number $M_{RR}$ of R-R computed
moments.\bigskip\ This possible disagreement is, however, of a logistic
rather than fundamental nature.

The required number of expansion coefficients $b_{i}$ in Eq. (24) is of
interest. \ At equilibrium only one is needed ($b_{0})$, as that situation
is described by a grand-canonical distribution function that is Gaussian. \
Next, if the system is sufficiently close to equilibrium then very few are
needed. \ Hence, near-equilbrium cases should pose little numerical
difficulty.

Summing up, the approach given in this paper will give exactly the same
solutions {\it at the fixed (but arbitrary) time }$t{\it \ \ }${\it as does
the R-R approach}. Therefore, for fixed $n,$\ our $c_{in}(t)$s coincide with
the R-R $a_{i}(t)$s and our $p(x|t)$ coincide with the R-R $f(x,t)$. This
holds at each time $t$ (Cf. Eq. (8)). For any other time value, $t^{\prime }$%
, say, we would have to input new $<A_{k}>$ values appropriate for that
time. R-R, instead, get coefficients $a_{i}(t)$ valid for continuous time $t,
$ since they are using Boltzmann's transport equation, which is a continuous
one. Our approach, by contrast, yields solutions valid at discrete point of
time $t$. This distinction, ``discrete versus continuous'', does not
compromise the validity of the Fisher- Schroedinger, non-equilibrium
thermodynamics bridge that we have built up here.

\bigskip

\section{Conclusions and Discussion}

It is becoming increasingly evident \cite%
{f0,f1,f2,f3,f7,pla1,pla2,pla5,roybook} that Fisher information $I$ is vital
to the fundamental nature of physics. In a previous effort \cite{nosotros},
we showed how the $I$-concept lays the foundation for a thermodynamics {\it %
in the usual equilibrium }case. Here we have shown that $non-equilibrium$
thermodynamics case can likewise be formed in this way.\ This considerably
expands the horizon envisioned in \cite{nosotros}

The main result of this work is the establishment, by means of Fisher
information, of a connection between non-equilibrium thermodynamics and
quantum mechanics. The emphasis here lies in the word ``connection''. Why
would such a link be of interest? Because it clearly shows that
thermodynamics and quantum mechanics can both be expressed by a \ formal SWE
(20), out of a common informational basis [21].\

The physical meaning of this SWE is flexible, since its ''potential
function'' $U(x)$ originates in data $<A_{k}>_{t},$via Eq. (21), of a $%
physically$ $general$ nature. \ This depends upon the application. \ The $%
<A_{k}>_{t}$ are introduced into the theory as $empirical$ inputs. The
approach also encompasses quantum effects. \ In the latter cases the
effective potential function includes quantum effects.\ Also, the Planck
constant $\hbar $, which does not explicitly appear in Eq. (20), would
appear in one or more inputs $<A_{k}>_{t}$as, for example, would occur if
the expectation value of the linear momentum of an electron were measured. \
The classical Boltzmann equation of the R-R approach would then of course $%
not$ be useable. \ In this way, our approach encompasses both quantum- and
classical thermodynamic effects.

\bigskip

\bigskip



\newpage

\begin{references}
\bibitem{katz} E. T. Jaynes in {\it Statistical Physics}, ed. W. K. Ford
(Benjamin, New York, 1963); A. Katz, {\it Statistical Mechanics}, (Freeman,
San Francisco, 1967).

\bibitem{nosotros} R. Frieden, A. R. Plastino, A. Plastino, and B. H.
Soffer, Phys. Rev. E {\bf 60, } 48 (1999).

\bibitem{fisher} R. A. Fisher, {\it Proc. Camb. Soc.} {\bf 22,} 700 (1925).

\bibitem{f0} B. R. Frieden, {\it Am. J. Phys. } {\bf 57}, 1004 (1989).

\bibitem{f1} B. R. Frieden, {\it Phys. Lett. A} {\bf 169}, 123 (1992).

\bibitem{f2} B. R Frieden, in: Advances in imaging and electron physics Vol.
90, ed. P.W. Hawkes (Academic, New York, 1994) pp. 123-204.

\bibitem{f3} B. R. Frieden, {\it Physica A} {\bf 198}, 262 (1993).

\bibitem{f4} B. R. Frieden and R. J. Hughes, {\it Phys. Rev. E} {\bf 49},
2644 (1994).

\bibitem{f5} B. Nikolov and B. R. Frieden, {\it Phys. Rev. E} {\bf 49}, 4815
(1994).

\bibitem{f6} B. R. Frieden, {\it Phys. Rev. A} {\bf 41}, 4265 (1990).

\bibitem{f7} B. R. Frieden and B. H. Soffer, {\it Phys. Rev. E} {\bf 52},
2274 (1995).


\bibitem{f8} B. R. Frieden, {\it Founds. Phys.} {\bf 21}, 757 (1991).

\bibitem{silver} R. N. Silver in {\it E. T. Jaynes: Physics and Probability}%
, edited by W. T. Grandy, Jr. and P. W. Milonni (Cambridge University Press,
Cambridge, 1992).

\bibitem{pla1} A. Plastino, A. R. Plastino, H. G. Miller, and F. C. Khana,
{\it Phys. Lett. A} {\bf 221}, 29 (1996).

\bibitem{pla2} A. R. Plastino and A. Plastino, {\it Phys. Rev. E} {\bf 54},
4423 (1996).

\bibitem{pla3} A. R. Plastino, A. Plastino, and H. G. Miller, {\it Phys.
Rev. E} {\bf 56}, 3927 (1997).

\bibitem{pla4} A. Plastino, A. R. Plastino, and H. G. Miller, {\it Phys.
Lett. A} {\bf 235}, 129 (1997).

\bibitem{ku} S. Kullback, {\it Information theory and statistics} (Wiley,
NY, 1959).

\bibitem{ravi} M. Ravicule, M. Casas, and A. Plastino, {\it Phys. Rev. A}
{\bf 55} (1997) 1695.

\bibitem{pla5} A. R. Plastino, M. Casas, and A. Plastino, {\it Phys. Lett. A}
{\bf 246}, 498 (1998).

\bibitem{roybook} R. Frieden, {\it Physics from Fisher information}
(Cambridge University Press, Cambridge, England, 1998).

\bibitem{paul} P. I. Richards, {\it Manual of mathematical physics}
(Pergamon Press, London, 1959) p. 342.




\bibitem{librobis} Y. B. Rumer and M. S. Ryvkin, {\it Thermodynamics,
statistical mechanics and kinetics} (MIR Publishers, Moscow, 1980).

\bibitem{} B. R. Frieden and B. H. Soffer, ${\em Found.of}$ ${\em Physics}$%
{\em \ }{\bf 13}, 89 (2000).
\end{references}
\end{document}